# Research on a Hybrid System With Perfect Forward Secrecy

Weiqing You[1], Guozhen Shi[1], Xiaoming Chen[1], Jian Qi[1], Chuang Qing[1]

1. Beijing Electronic Science & Technology Institute, Beijing, China

winw2012@163.com, 894572560@qq.com

**Abstract**—The rapid development of computer technology will be the whole world as a whole, the widespread application of instant messaging technology to bring great convenience to people's lives, while privacy protection has become a more significant problem. For ordinary it's hard to equip themselves with a cryptograph machine. In this paper, through in-depth study of elliptic curve cryptosystem ECC and advanced encryption standard AES encryption algorithm, according to the characteristics of public key cryptography, elliptic curve version through the establishment of Diffie-Hellman key exchange protocol, combined with AES, design a set of perfect forward secrecy mixed cryptograph system. The system can guarantee the security of communication, easy to implement, the operation speed is quick and the cost is low. At last, the security of the system is analyzed under the environment of common network attacks.

## I. INTRODUCTION

With the rapid development of information technology and Internet technology, communication technology service system and application booming, brings great convenience to people's lives, but also brought new challenges, the security problem in network communication has attracted more and more attention. At present, how can the communication security in the open network environment be guaranteed in network communication?. With the advent of the information age, information security technology has also become increasingly important. Data security and complete implementation of online transmission are the research purposes of information security technology. Among them, data encryption technology is the core of information security technology. Data encryption as a data security technology has a long history, dating back to ancient times, such as the 2000 BC, the ancient Egyptians will use some methods to protect their written information. Coding the information has been used by Julian Caesar (Kaiser) and used in all wars, including the American Revolution, the American Civil War, and the two world war. The most well known is the World War II (1942) in Germany used to encrypt information encoding machine German Enigma (British Nigema) machine.

In the past, the cryptograph technology only for military and defense, is now the new network of space is inevitable to affect the life of each individual in the network space and the virtual space is easily attacked, in the big data and cloud computing is highly developed today, people's privacy has been a great challenge. Although Commercial Cryptography machines have been put into use, there is still no suitable, cheap and convenient cryptograph product for individuals to protect their privacy. Events such as WeChat chats and QQ chats are exposed, and encrypted mail exposure is endless. This paper aims to now want to study sophisticated encryption technology, advanced encryption standard AES[1] and the characteristic of ECC[2] public key cryptographic algorithm based on elliptic curve, establish a set of security available mixed encryption system, and provides a design scheme of cryptographic system. The use of AES encryption technology can satisfy even the amount of data to be encrypted, you can also quickly complete encryption, AES encryption plaintext, the key used to ECC to encrypt.

ECC algorithm is more favored in communication network authentication, [3] and key distribution [4] scheme. It was proposed by N.Kobliz and V.Miller in 1985 that [2] Curve Cryptography (ECC) is applied to the public key cryptography system, and it is the focus of Elliptic research at home and abroad. An Elliptic Curve Cryptosystem with a key length of 160 bits, and its security is equivalent to a 1024 bit RSA[5] system. At present, it has been widely used in smart cards[11], e-commerce and other business systems. ECC generates private keys, public keys, and saves memory space, bandwidth, and processing time.

Because the mixed encryption algorithm, which takes advantage of public key cryptography and security advantage in key management and distribution, and use of the advantages of symmetric cryptosystem speed, so it is widely used in the field of information.

## II. THE BASIC THEORY OF CRYPTOGRAPHY DESIGN

### A. Overview of Cryptography

The encrypted information is called plaintext, and the encrypted information is called ciphertext, and the transformation from plaintext to ciphertext is called encryption transformation. On the contrary, the transformation from ciphertext to plaintext is called decryption transformation. A key is a parameter that controls the encryption, transformation, and decryption transformations.

The key is determined, and the encryption and decryption transformation is determined. A cryptosystem is usually made up of the above five parts[6,10], note that $<M, C, K, \varepsilon, D>$:

M is a collection of all plaintexts while C is the collection of all ciphertexts. K is a collection of keys, usually each key k,k=<ke,kd> and consisting of the encryption key Ke and the decryption key $K_d$, which may be equal to $K_d$ and $k_e$. $\varepsilon$ and $D$ represent the set of encryption algorithms and the set of decryption algorithms, respectively.

If the encryption key of a cipher body is the same as the decryption key, it is called symmetric cryptosystem, otherwise it is called the public key cryptosystem, and symmetric cipher and public key cipher have their advantages and disadvantages. When designing cryptographic systems, they can be used in combination to achieve optimal results.

*B. Ecc and AES*

Menezes-Vanstone[2] public key cryptosystem is an efficient implementation of ElGamal[7] public key cryptosystem in elliptic curves(Abbreviation ECC). It was proposed by A.J.Menezes and S.A.Vanstone in 1993 and is still regarded as a secure crypto-system. If you want to know more about it, please refer to the literature[2]. Public key cryptography on elliptic curves can be used not only for public key cryptography, but also for key agreement. It has high security and is widely used in various security fields.

Scholars in the field of information security must be very familiar with AES, which is characterized by fast computing speed, and can carry out a large number of data encryption and transmission. Although it is not as secure as ECC, it is obvious that AES is more appropriate when dealing with huge amounts of data. For more details, please refer to the literature[1].

### III. KEY ISSUES IN THE DESIGN OF CRYPTOGRAPHIC SYSTEMS

*A. The Demand Analysis*

In the rapid development of information technology in today's society, privacy protection has become a very serious problem. WeChat, Tencent, Facebook and other instant messaging tools widely into people's lives, these tools are easy to use, convenient for people's life greatly. All have huge information interaction, at the same time, this information is exposed to a very insecure network environment, with the rapid development of big data and cloud computing, people through the use of these communication tools has greatly exposed their privacy. There is no denying that all the information transmitted through these tools is stored without reservation on the server developing these tools.

Therefore, having a cheap and efficient privacy protection software is of great significance in the era of big data. Through the design of an encryption software, due to the use of a large number of data encryption, this requires fast enough encryption speed, high efficiency, and easy to implement. Advanced Encryption Standard AES is very good at a large amount of data encryption and decryption, but its key distribution is a big problem, Diffie-Hellma[8] key exchange protocol is a good solution to this problem, public key cryptography can be used to protect the key, because the operation overhead is not conducive to a large number of data encryption and decryption. It is a very worthwhile practice to mix the public key cryptography algorithm with the symmetric key algorithm according to certain rules. This design, through the public key algorithm and the private key algorithm research, designs a security, the convenient cryptograph system, can very effectively protect the privacy.

Through the installation of the cryptograph system on the PC, through a friendly interface to provide users with confidentiality services.

*B. Random Number Module Design*

Random number module is the key module in all cryptographic systems. In literature [2 and 8], the random number is used to generate key and secret index needed by algorithm.

The best approach may be to use the concept of entropy [9], according to the national cryptograph Administration issued by the GM/T 0005 "random number testing specification" related requirements, produce a random number in line with product requirements. Of course, when we use software to implement the system, the random number quality may not have any random number, the random number generated by the chip is random.

This paper proposes using AES[1] algorithm to output the ciphertext as random number. The security of AES guarantees the quality of random numbers, and it has computational advantages when we implement it with software.

*C. Key Management Module Design*

Key management includes key generation, distribution, preservation, modification, query and so on. Among them, the generation and preservation of keys may be the most difficult problem.

A key function of public key cryptography is to solve the key distribution problem. Most hybrid encryption systems usually use digital certificates to ensure authentication security, and then use public key technology to realize the transmission of keys. However, the hybrid system designed in this paper is willing to choose another way of using ECC and dig it out. The hybrid system does not use the ECC public key encryption algorithm, but designs a ECC version of Diffie Hellman[8] key exchange system. In this system, the main task of the key management module is to secure a symmetric key that is shared by both sides for a long period of time. This is undoubtedly a good thing for the design of the key management module, because it is simple and simple, which can make the key management easier and make the system more secure.

*D. Parameter Setting*

According to the relevant standard of SM2 elliptic curve public key cryptography issued by the national cipher administration, the correlation coefficients are designed.

*E. Cryptographic Service Interface*

The system provides two interfaces: an internal parameter setting interface and an application service interface.

The internal parameter setting interface is used to generate random numbers, calculate session keys, load long-term keys, etc. the application service interface provides several loading items, and the user can select the application services that need to be protected. For example, when passing messages on the QQ, first read the plaintext message through the background, then encrypt it locally and send it to the other party of the communication, and the Tencent server stores only encrypted ciphertext of local messages.

## IV. A HYBRID SYSTEM WITH COMPLETELY FORWARD SECRECY

Through the study of the pros and cons of ECC and AES, this paper proposes a cryptograph design scheme based on ECC and AES, which is divided into three parts: identity authentication, session key establishment and secret communication.

### A. Two Party Communication Algorithm

In order to describe the convenience, Alice and Bob are still used to represent the two communications, and before that, they already have a long-term shared key K. In each communication, the system uses its random number module to generate secret index A and B for the user, P is one of the points on the elliptic curve, and P is open. Specific programs are as follows:

**Step1:** *Alice* uses the random number module to generate the secret digital *A* and calculates $AES(AP, K)$ and sends it to Bob;

**Step2:** *Bob* uses the random number module to generate the secret digital B, calculates $AES(BP, K)$ and sends it to Alice;

**Step3:** *Alice* uses its shared key *K* to compute $AES^{-1}(AES(BP, K))$, and gets *BP*. Since *Alice* has a secret number *A*, it can calculate $K_S = ABP$, then destroy the secret number *A*;

**Step4:** *Bob* uses its shared key *K* to compute $AES^{-1}(AES(AP, K))$ *Bob* get *AP*, because Bob has secret digital *B*, so it can calculate $K_S = ABP$, then destroy secret digital *B*;

**Step5:** *Alice* and *Bob* access to the communication session key $K_s$, then the two sides using session key $K_s$, using *AES* algorithm for data transmission. That is:

*Alice* send $c_1 = AES(m_1, K_s)$ to *Bob* while *Bob* send $c_2 = AES(m_2, K_s)$. So Both of *Alice* and *Bob* can carry out the decryption through $K_s$.

The first and second steps establish the authentication mechanism for the hybrid system, and the latter will analyze the authentication mechanism, which is completely positive and confidential. The third and fourth steps are calculated simultaneously by *Alice* and *Bob*. It is important to note that in the third and fourth steps, it is necessary to destroy the secret index *A* and *B*.

The user may still have questions about the design in multiparty secure communications, and the answer is yes. The design above does provide a secure two party communications design, but that does not mean that the design can only be limited to communications between the two parties. In depth research, it is not difficult to find out how many people are on the network side. We can always share the key in a particular group beforehand, and the shared key can be completed whether it belongs to the certification work of this particular group. The problem with the system, then, is the problem of attacks within trusted groups.

### B. Multiparty Secure Communication Algortihm

Since it is for a particular group, the problem will be simple. Perhaps the most direct way is for each individual in the group to have a unique identity symbol. Public key technology will be used here. In a trusted LAN, a public private key pair is generated for each user, and each person's public key is published, and the private key is handed to each user. $K_{Ap}$ and $K_{Bp}$ are the private keys of Alice and Bob, respectively. $K_A$ and $K_B$ are the keys of *Alice* and *Bob* respectively, and *H* is the hash function preset in the system. This system adopts *MD*5. Now, we propose multiparty secure communication systems:

**Step1:** *Alice* generates a secret number *A* using the random number module. Calculate:

$$T = AES(AP || ECC_{sig}[h(AP), K_{AP}]; K)$$

and send *T* to Bob ( $A || B$ indicates attaching the message *A* to the B ).

**Step2:** *Bob* uses the random number module to generate secret numbers *B*. Calculate

$$S = AES(BP || ECC_{sig}[h(BP), K_{AP}]; K)$$

and send *S* to *Alice*.

**Step3:** *Alice* decrypts the received message by using its shared key, obtaining *BP*, $ECC_{sig}[h(BP), K_{Bp}]$. By using the *Bob's* public key to verify the correctness of the signature algorithm, the hash value is calculated to ensure that the data has not been modified. These steps can effectively confirm the identity of *Bob*. Since Alice has secret digital *A*, it is possible to compute $Ks = ABP$, and destory the *A*;

**Step4:** *Bob* decrypts the received message by using its shared key, obtaining *AP*, $ECC_{sig}[h(AP), K_{Ap}]$.

Then the *Alice* public key is used to verify the signature algorithm and the hash value of the *AP* is calculated to ensure that it is not modified by string. These steps can effectively confirm the identity of *Bob*. Since Bob has secret digital *B*, it is possible to compute $Ks = ABP$, and destroy the *B*;

**Step5:** *Alice* and *Bob* have access to the communication session key *K*s, using the *AES* system, they can communicate with each other confidential.

The, we can see the system is perfect forward secrecy system above. Because of the limitation of space, let's talk about the system how work, so the security analysis of the hybrid system needs to be postponed for the moment.

## V. SYSTEM FUNCTION DESIGN

### A. Load global key K

The long time shared key $K$ is generated by using the random number generating module, and the key file is passed to the other party of the communication through the secure channel.

### B. User authentication and session key generation design

The user $A$ software startup, and generate a secret digital operation to generate the intermediate variables of $ECC$ based key exchange, take the first signature encryption, signature $ECC$ algorithm, encryption algorithm is $AES$ algorithm, the final encryption result to the user $B$; $B$ users after receiving the generated secret numbers and exchange operations to generate intermediate variables ECC based key, take the same signature then encryption way of computing results will be sent to $A$. User $A$ and $B$ complete the authentication process and calculate the session key of the communication by their secret number. When the session key is obtained, the encrypted message is encoded and encrypted according to the plaintext entered by the system.

### C. System Fault Design

In the course of communication, there may be a disconnection. When the completion of key agreement, the two sides began communication, because of the failures of the communication of the party may appear disconnected, start recording 30 seconds short time, system connection will use the last key to restore the line to accept the ciphertext, more than 30 seconds when the short-term need re key agreement, to accept information consultation and prompts the other off-line restart state negotiation after the completion of the new key to replace the old key. After ten minutes of disconnection, the system automatically destroys the session key.

## VI. SECURITY ANALYSIS

### A. Intermediate Attack Analysis

The hybrid system uses the Diffie-Hellman[8] key exchange protocol based on the elliptic curve version to obtain the session key for each communication, then the most worrying is probably the middleman attack. Here, we put aside the cost of digital certificates. In fact, digital certificates require that you have to trust an organization, but the design of this article does not need to consider this, and it also reduces costs for system design. Since the system is designed to protect privacy, it is feasible to share a long-term key through a secure channel for a small group of people. Here, we assume that the long key $K$ is secure, and then the key negotiation process can be effectively encrypted by a long key to prevent the middleman attack. Very cautious users may not be worried about whether the system might be replayed. The system does not add timestamp because it may be vulnerable to changes in the system within the group. But this is not to cause too much to worry about, in the design of the first two steps we are very determined to complete the whole process of identity authen-tication, key agreement in the protection of security key K, only in communication has no secret secret number, unable to establish a digital key agreement is also difficult to obtain, replay attack a successful attack on the system.

### B. Forward Search Attack Analysis

The biggest role of long-term key $K$ is to ensure that users involved in the communication can correctly identify each other's identity. Long term key $K$ may be the easiest to cause an attack. Since the key K is shared through secure channels, a smart attacker will not want to break the $AES$ cryptosystem [1], so he might think of forward search attacks. But note here that it's not easy to start a forward search attack on a relatively closed system, and it's more off the hook. For the forward search attack, the system user can safely use it. In fact, the hybrid system provides a fully forward secrecy mechanism.

The security of the hybrid cryptosystem allows attackers Turkey a hard time to start, but smart Turkey hides in the dark and records all the ciphertext that is being transmitted by the system. Assume that during long communications, the attacker Turkey captures a certain opportunity and can access the long key $K$. In this system, the Turkey obtains the key K and cannot decrypt the ciphertext it has recorded in the past. Because every communication, have carried out a key agreement, Turkey $K$ has the key to the history of every communication in key negotiation process decryption, but users in every key agreement after the completion of the destruction of the secret number, no secret number, even if the user is difficult to decrypt the ciphertext (I believe very few some people can remember a random string of more than 128bit), and Turkey. So even if Turkey gets the key, $K$ can't decrypt the ciphertext message of the past. However, when Turkey gets a long-term key, it can act as an intermediary attack.

In addition, the key K may be compromised in a complex network environment. When the adversary obtains the key K, it can obtain the interactive data: AP, BP. It is worth noting that, through AP, BP can not find ABP[13].

## VII. CONCLUSIONS

In this paper, a two-way mutual authentication protocol is designed. The authentication process and the key agreement process are completed simultaneously and have the function of the $PFS$[12] system. The DH key exchange protocol [8] based on $ECC$ uses the private key to sign the key negotiation process, and uses the long-term key encryption to generate the session key of each call. The session key is loaded into the AES algorithm system and has high encryption and decryption functions.

The system design details are not perfect, and there are still many problems to consider in the specific engineering applications. The difference between the Android version and the PC version, as well as the magnitude of the design requirements, etc., is a question worthy of further study.

ACKNOWLEDGMENT

This work was supported by National key R & D plan under grand: No.2016YFB0800304 and Beijing Natural Science Foundation of China:No.4102056.